\documentclass[prl,twocolumn,twoside,a4paper,showpacs,showkeys,amsmath,amsfonts]{revtex4} 

\usepackage[latin1]{inputenc}
\usepackage[T1]{fontenc}

\usepackage{pxfonts} %e-print

\usepackage{hyperref}

\newcommand{\preprintonly}[2][]%
                             {#2}%for the preprint 
\newcommand{\arxiv}[1]{arXiv \href{http://arXiv.org/abs/#1}{#1}}
\newcommand{\aarxiv}[1]{\preprintonly{, \arxiv{#1}}}
\newcommand{\cboxed}[1]{\preprintonly[#1]{\boxed{#1}}}
\newcommand{\ie}{\emph{i.e.}}
\newcommand{\etal}{\emph{et al.}}
\newcommand{\prletal}[1]{\preprintonly[ \etal]{#1}}

\newcommand{\avrg}[2][]{\langle #2 \rangle_{#1}}
\newcommand{\V}[1]{V_{#1}}

\newcommand{\Vc}[2]{V_{#1\vert#2}}
\newcommand{\Cor}[2]{\avrg{#1 #2}}
\newcommand{\e}{\mathrm{e}}

\newcommand{\fntA}[1]{{\mathsf{#1}}}
\newcommand{\A}{\fntA{A}}
\newcommand{\X}{\fntA{X}}
\newcommand{\B}{\fntA{B}}
\newcommand{\Q}[1][]{Q_{#1}}
\newcommand{\QB}{{\Q[\B]}}
\renewcommand{\P}[1][]{P_{#1}}
\newcommand{\PB}{{\P[\B]}}
\newcommand{\Y}{\fntA{Y}}
\newcommand{\QY}{{\Q[\Y]}}
\newcommand{\PY}{{\P[\Y]}}
\newcommand{\E}{\fntA{E}}
\newcommand{\N}{\mathrm{noise}}

\newcommand{\dDI}{\Delta I_{\mathord{\vartriangleright}}}
\newcommand{\rDI}{\Delta I_{\mathord{\vartriangleleft}}}

\newcommand{\ent}[1]{H_{#1}}
\newcommand{\entC}[2]{\ent{#1\vert#2}}
\newcommand{\I}[2]{I_{#1;#2}}
\newcommand{\bO}[1]{\mathcal{O}\big(#1\big)}

\newcommand{\Trs}{T}%{\eta}

\newcommand{\AB}{{\A\B}}
\newcommand{\n}{n}
\newcommand{\coll}{{\mathrm{coll}}}
\newcommand{\coh}{\coll}
\newcommand{\het}{{\mathrm{het}}}
\renewcommand{\hom}{{\mathrm{hom}}}

\newcommand{\f}[1]{\tfrac1{#1}}

\begin{document}
\title{Collective attacks and unconditional security \\
  in continuous variable quantum key distribution} 
\date{\today}

\author{Frédéric \surname{Grosshans}} 
\email[Permanent e-mail: ]{frederic.grosshans@m4x.org}
%\altaffiliation[Current address: ]{Somewhere}
\affiliation{Max-Planck-Institut für Quantumoptik,
  Hans-Kopfermann-Str. 1, D--85746 Garching, Germany}


\begin{abstract} 
  We present here an information theoretic study of Gaussian
  collective attacks on the continuous variable key distribution
  protocols based on Gaussian modulation of coherent states \cite{GG,
    GG-Proc, hetero}. These attacks, overlooked in previous security
  studies, give a finite advantage to the eavesdropper in the
  experimentally relevant lossy channel, but are not powerful enough
  to reduce the range of the reverse reconciliation protocols. Secret
  key rates are given for the ideal case where Bob performs optimal
  collective measurements, as well as for the realistic cases where he
  performs homodyne \cite{GG, GG-Proc} or heterodyne measurements
  \cite{hetero}. We also apply the generic security proof of
  Christiandl \etal\ \cite{generic} to obtain unconditionally secure
  rates for these protocols.
\end{abstract}
\maketitle

Over the past few years, quantum continuous variables (CV) have been
explored as an alternative to qubits for quantum key distribution
(QKD) \cite{hillery, gottesman, CLVA}. More specifically, protocols
using coherent states and homodyne \cite{GG, GG-Proc, silberhorn} or
heterodyne \cite{hetero} measurements have been proposed and
experimentally demonstrated \cite{nature, lorenz}. Relying on
technologies allowing much higher rates than allowed by the single
photon detectors used in qubit based QKD, those protocols are the
only ones which could allow key rates in the GHz range in the
foreseeable future.

However, the security proofs of these new protocols are not yet as
strong as the ones of the qubits-based protocols: they are almost all
limited to individual or finite-size \cite{GC} attacks. To our
knowledge, the only unconditional security proofs of CV QKD protocols
are \cite{gottesman, barcelona, IVAC}, only the latter studying
coherent-states based protocols. These proofs all rely on specific
suboptimal key extraction procedures, and, for each case, it's
difficult to separate the effects of the technical inefficiencies 
of the encoding scheme from the more fundamental effect of real attacks
possibility of Eve --- the eavesdropper.
 
In this letter, we study the effects of a Gaussian collective attack
on the key rate of CV QKD protocols based on the Gaussian modulation
of coherent-states \cite{GG, GG-Proc, hetero} sent through a lossy
channel. In these attacks, Eve uses a Gaussian unitary to interact
with each of the transmitted pulse and stores her ancillas into a
quantum memory. She performs then a collective measurement on her
ancillas after Alice and Bob --- the partners performing QKD --- have
used the public classical channel to fulfill the protocol.

After introducing the notations used in this letter, we recall the
values of various information theoretic quantities for Gaussian
states. Then, we compute the secret key rate which can be achieved
using direct reconciliation when Bob is allowed to do collective,
heterodyne or homodyne measurements. Those results are then extended
to reverse reconciliation protocols and compared with the
unconditionally secure rates obtained from the generic security proof
of Christiandl \etal\ \cite{generic}.

While completing this work, we learned Navascués and Acín used very similar
techniques to study the security bounds of these protocols
%had
%obtained simultaneously and independently very similar results
\cite{NA}.

\paragraph{Notations.}

In all the QKD protocols discussed in this letter, Alice sends $\n$
Gaussian modulated coherent states through a lossy channel of
transmission $\Trs$. 
%In principle, the coherent state could be
%generated with a pair of entangled beams, one of them being measured
%with an heterodyne detection by Alice \cite{virtual}. This equivalent
%entanglement-based protocol will sometimes be useful in the following. 

Bob then makes measurements on the pulses he receives. It can be an
optimal collective measurement or, more realistically, a series of
heterodyne \cite{hetero} or homodyne \cite{GG, GG-Proc, nature}
measurements. Alice and Bob then agree on a secret key through a
(direct or reverse) reconciliation procedure. We are interested in the
asymptotic key rates obtained at the limit $\n\to\infty$.

In the following, $\A$ refers to the quantum state of the light pulse
prepared by Alice, $\B$ and $\E$ to the one received by Bob and Eve.
$\X$ refers to the (classical) value of Alice's modulation and $\Y$ to
the one of Bob's measurement. For instance $\ent{\B}$ will denote the
Von Neumann entropy of the density matrix $\rho_\B$ at Bob's side,
while $\ent{\Y}$ will denote the Shannon differential entropy of Bob's
measurements.

Alice modulates the two quadratures $\Q[\X]$ and $\P[\X]$ of the
coherent states she sends with random values following Gaussian
distribution. To simplify the analysis, we will assume this modulation
to be symmetric in $\Q$ and $\P$.  

If Bob performs a heterodyne measurement, he gets the two noisy
measurements $\Q[\Y]^\het=\Q[\B]+\Q[\N]$ and
$\P[\Y]^\het=\P[\B]+\P[\N]$, where $\Q[\N]$ and $\P[\N]$ are two
independent Gaussian random variables of variance 1. (The units used
in this letter correspond to a unity variance of the vacuum.)  If he
performs a homodyne measurement, he perfectly measures one quadrature
--- let say $\Q$ --- and obtains no information on the other --- $P$.
One has therefore $\Q[\Y]^\hom=\Q[\B]$ and $\P[\Y]^\hom=\P[\N]$ (or,
of course, the symmetric case, where $\P[\Y]^\hom=\P[\B]$ and
$\Q[\Y]^\hom=\Q[\N]$).

The lossy channel is modeled by a beamsplitter of transmitivity
$\Trs$ and reflectivity $1-\Trs$, the reflected beam being given to
Eve.  In the equivalent entanglement-based scheme \cite{virtual}, this
attack gives Eve the purification of the mixed state
$\rho_\AB^{\otimes\n}$ shared by Alice and Bob. Eve then performs a
collective measurement on her part of the purification, after Alice
and Bob's classical communication has occurred.  As shown below, this
attack is more powerful than the ones studied in \cite{GG, GG-Proc,
  nature, GC, hetero}.  However, this attack model is not generic in
two aspects.

First, the channel model itself is not generic, since we restrict
ourselves to the lossy channel and omit to consider nonzero
added-noise and non-Gaussian attacks. This restriction is only due to
brevity consideration and will be lifted in a longer article
\cite{long}, which will also contain a study of squeezed states
protocols. Of course, in an experiment, the amount of added noise has
to be measured by Alice and Bob through some sampling and will never
be exactly zero.  They would thus have to use the more general results
of \cite{long}.

A more fundamental restriction comes from the fact that 
%we restrict ourself to collective attacks, \ie\ 
we suppose Alice and Bob share a
state of the form $\rho_\AB^{\otimes\n}$. In other words, we restrict
Eve to individual attacks on the channel, even if she is allowed to
make collective measurements on the ancillas obtained through these
attacks. This restriction will be lifted at the end of this letter,
where we apply the generic security proof \cite{generic} which does
not rely on any assumption about Eve's attack.

\paragraph{Entropies and mutual information.}

Let $\V{\QB}$ ($\V{\PB}$) be the variance in the $\Q$-quadrature (in
the $P$-quadrature) of the Gaussian state $\rho_\B$. Since squeezing
is a reversible operation, it doesn't alter the Von Neumann entropy of
$\rho_\B$, which is an increasing function of $\V\B:=\sqrt{\V\QB \V\PB}$
\cite{entth}:
%, where the variance of the vacuum is set to unity, 
\begin{equation}
  \begin{split}
    \ent{\B} & =\tfrac{\V\B+1}{2}\log\tfrac{\V\B+1}{2}
    -\tfrac{\V\B-1}{2}\log\tfrac{\V\B-1}{2}
    \\
    &=\log\tfrac{\V\B+1}2
    +\tfrac{\V\B-1}2\log\tfrac{1+1/\V\B}{1-1/\V\B} .\label{eq:HB}
  \end{split}
\end{equation}
The logarithms in the above expression should be taken in base 2 if one
wants the result in bits, or in base $\e$ if one wants it in nats.
For strong modulation ($\V\B\gg 1$), one will use the Taylor expansion 
\begin{equation*}
  \ent\B=\log\V\B +\log\tfrac{\e}2 +\bO{\tfrac1{\V\B}}.
\end{equation*}

Let $\V\QY$ and $\V\PY$ be the variances of Bob's two orthogonal
quadrature measurements. The Shannon differential entropy $\ent\Y$ is simply
the logarithm of $\V\Y:=\sqrt{\V\QY \V\PY}$, up to an
arbitrary additive constant \cite{Shannon}, which can be set to 0:
\begin{equation}
  \label{eq:HY}
  \ent{\Y}
  =\log V_\Y.  
\end{equation}

%\paragraph{Mutual informations.} 

The rate of common information Alice and Bob can extract from their
classical values is given by the mutual information \cite{Shannon}
\begin{equation*}
  \I\X\Y:=\ent\Y - \entC\Y\X=\ent\X+\ent\Y-\ent{\X\Y}.
\end{equation*}
If Bob uses a heterodyne detection, which adds a unit of noise,
$\V\Y^\het=\V\B+1$.  Since a coherent state sent through a lossy
channel stays a coherent state, the conditional variance
are $\Vc\B\X=1$ and $\Vc\Y\X^\het=\Vc\B\X+1=2$. One has therefore 
\begin{equation}
  \label{eq:IXYhet}
  \I\X\Y^\het=\log\tfrac{\V\B +1}{2}.
\end{equation}

If Bob is allowed to make arbitrary  measurements on the pulses, the
information they share is then given by the Holevo information  \cite{H, SW}
\begin{equation*}
  \I\X\B:=\ent\B - \entC\B\X,
\end{equation*}
which is attained by collective measurements.
Since Bob receives pure (coherent) states, $\entC\B\X=0$ and
\begin{equation}
  \label{eq:IXB}
  \I\X\B=\ent\B=\I\X\Y^\het+\tfrac{\V\B-1}2\log\tfrac{1+1/\V\B}{1-1/\V\B}.
\end{equation}
If $\V\B\gg1$, one has
\begin{equation}
    \label{eq:IXYIXB}
    \I\X\B=\I\X\Y^\het+ \log\e + \bO{\f{\V\B}}.
\end{equation}
Therefore, by using heterodyne detection instead of the optimal collective
measurement, Bob loses an amount of information up to $\log \e$ (\ie\ 
1~nat $\simeq$ 1.44~bits) per pulse.

\paragraph{Direct key distribution.}

To attain the rate $\I\X\Y$, Alice and Bob can use random codes of
size $\exp(\n\I\X\Y)$, where the basis of the logarithms and the
exponential are the same. Devetak and Winter have recently shown
\cite{DW} %\cite{DWshort, DWlong} 
that Alice can divide this code into privacy
amplification subsets of size  close to  $\exp(n\I\X\E)$.
This allows Alice and Bob to generate a secret
key through a direct reconciliation procedure using only forward
communication. This key can be generated at a rate asymptotically
close to
\begin{equation}
  \label{eq:dDIXYE}
  \dDI:=\I\X\Y-\I\X\E.
\end{equation}

If Bob makes the optimal collective measurement, substituting
$\I\X\Y^\coll=\I\X\B$ in the above expression and using equation
\eqref{eq:IXB} (with $\V\B=\Trs\V\A+1-\Trs$ and
$\V\E=(1-\Trs)\V\A+\Trs$) gives us the attainable direct
reconciliation key rate. At the high modulation limit, where
$\V\A\gg1/\Trs ; 1/(1-\Trs)$, one has \cite{noteapprox}
\begin{equation*}
  \cboxed{
    \dDI^\coll=\log\tfrac{\Trs}{1-\Trs}
    +\bO{\{\f\Trs+\f{1-\Trs}\}\f{\V\A}}.
    }
\end{equation*}
This limit is the same as the one found assuming Bob \emph{and Eve} are
restricted to heterodyne measurements \cite{hetero} (individual attacks).

\paragraph{Direct heterodyne key distribution.}

The equation \eqref{eq:dDIXYE} can be applied to the case where Bob
uses heterodyne detection \cite{hetero}. We have then a lossy channel
of transmission $\Trs$ between Alice and Bob and another lossy channel
of transmission $1-\Trs$ between Alice and Eve. We can therefore use
the equations \eqref{eq:IXYhet} and \eqref{eq:IXB} to expand this
expression into
\begin{equation*}
    \dDI^{\het} =\log\tfrac{\V\B+1}{\V\E+1}
    -\f{\V\E-1}2\log\tfrac{1+1/\V\E}{1-1/\V\E}.
\end{equation*}
%where $\V\B=\Trs\V\A+1-\Trs$ and $\V\E=(1-\Trs)\V\A+\Trs$. 
As shown by equation \eqref{eq:IXYIXB}, Eve can gain up to 1 nat per
pulse by using collective measurement. The best situation for Alice
and Bob is the high modulation limit $\V\A\gg\f{1-\Trs}$, where
\begin{equation*}
  \cboxed{
    \dDI^\het =\log\tfrac\Trs{1-\Trs}\f\e
    +\bO{\f{(1-\Trs)\V\A}}.
    }
\end{equation*}
It is therefore not possible to perform direct heterodyne QKD if the
channel transmission is smaller than $\Trs^\het_{\min} ={\e}/({\e+1})
\simeq0.73$. These maximal losses of 1.4~dB imply a shorter range for
this protocol than the 3~dB deduced if one only considers individual
attacks or if Bob uses optimal collective measurements.

\paragraph{Direct homodyne key distribution.}

Surprisingly, the original proposal \cite{GG} of direct homodyne QKD is
more robust.  For those protocols, Alice modulates both
quadratures $\Q$ and $\P$ with the same variance $\V\A$ and Bob
chooses randomly one quadrature to measure (let say $Q$). After the
public disclosure of this quadrature choice, the information on $\P$
is useless and can be forgotten by Alice. % \cite{homdirflow}.  
The state she has sent to Bob is therefore a mixture of coherent
states with a given value of $\Q$ but different values of $\P$. The
mixture received by Bob is a Gaussian mixed states with variances
$\Vc\QB\X^\hom=1$ and $\Vc\PB\X^\hom=\V\B$, therefore
$\Vc\B\X^\hom=\sqrt{\V\B}$. When $\sqrt{\V\B}\gg1$, one has
\cite{noteapprox}
\begin{align}
  \entC\B\X^\hom
  &=\log{\sqrt{\V\B}}+\log\tfrac\e2+\bO{\tfrac1{\sqrt{\V\B}}}
  \label{eq:HBXhom}
  \\
  \ent\B^\hom
  &=\log{\V\B}+\log\tfrac\e2+\bO{\tfrac1{\V\B}}
  \nonumber
  \\
  \I\X\B^\hom
  &=\f2\log\V\B+\bO{\f{\sqrt{\V\B}}}
  =  \I\X\Y^\hom+\bO{\f{\sqrt{\V\B}}}
  \nonumber.
\end{align}
Thus, collective measurements only give a small amount (of order
$1/\sqrt{\V\B}$) of supplementary information over homodyne detection,
the noise in the useless quadrature ($\P$) plays a crucial role in
this.

Eve receives similar mixed states, and, in the strong modulation regime 
($\V\A\gg1/\Trs; 1/(1-\Trs)$), one has
\begin{equation*}
  \cboxed{
    \dDI^\hom =\f2\log\tfrac{\Trs}{1-\Trs}
    +\bO{\big\{\f{\sqrt{\Trs}}+\f{\sqrt{1-\Trs}}\big\}\f{\sqrt{\V\A}}}.
  }
\end{equation*}
The advantage given to Eve by collective measurements is therefore of
order $1/\sqrt{\V\A}$ and can be arbitrarily reduced by Alice's use of
a strong enough modulation. Therefore, unlike the heterodyne protocol,
the key rate of the direct homodyne key distribution protocol remains
almost unchanged when compared to \cite{GG}, where Eve was restricted
to (post\-poned) homodyne measurements. More specifically the range
limit of this protocol stays at 3~dB (50~\%) of losses, whether one
considers collective measurements or not.

\paragraph{Reverse key generation.}

For symmetry reasons, backward communication is needed distribute a
quantum key beyond this 3~dB limit.  Either post\-selection
\cite{silberhorn, barcelona} or a reverse reconciliation procedure
\cite{GG-Proc, hetero, nature} can be used for this purpose.  In the latter
case, the attainable rate is given by \cite{DW}
%\cite{DWlong, DWshort}
\begin{equation*}
  \rDI:=\I\X\Y-\I\Y\E.
\end{equation*}

If Bob performs an optimal collective measurement, one has to replace
the above expression by $\I\X\B-\I\B\E$, where $\I\B\E$ is the
quantum mutual information
\begin{equation*}
  \I\B\E:=\ent\B + \ent\E - \ent{\B\E}.  
\end{equation*}
Since the joint state $\rho_{\B\E}$ is obtained by the (reversible)
mixture of $\rho_\A$ and a vacuum state in the mode $\fntA{N}$, one
has $\ent{\B\E}=\ent\A+\ent{\fntA{N}}=\ent\A$.  At the high modulation
limit ($\V\A\gg1/\Trs ; 1/({1-\Trs})$), one has therefore \cite{noteapprox}
\begin{gather*}
  \I\B\E=\log\V\A+\log\Trs(1-\Trs)\tfrac\e2
  +\bO{\{\f\Trs+\f{1-\Trs}\}\f{\V\A}}\\
  \cboxed{\rDI^\coh=\log\f{1-\Trs}+\bO{\{\f\Trs+\f{1-\Trs}\}\f{\V\A}}}, 
\end{gather*}
which is, like in the direct case, very close to the result obtained
with heterodyne detection in an individual attacks scenario
\cite{hetero}. For strong losses ($1/\V\A\ll\Trs\ll1$), this
expression becomes
\begin{equation*}
  \rDI^\coh=\Trs\log\e+\bO{\Trs^2+\f{\Trs\V\A}}. 
\end{equation*}

\paragraph{Reverse heterodyne key generation.}

In the heterodyne case \cite{hetero}, which is symmetric in $\Q$ and $\P$, one has
\begin{gather*}
%  \Vc\E\Y^\het=
  \Vc{\P[\E]}\Y^\het=\Vc{\Q[\E]}\Y^\het%\\
  :=\V{\Q[\E]} - \tfrac{\Cor{\Q[\E]}{\Q[\Y]^\het}^2}{\V{\Q[\Y]}^\het}%\\
%  &=(1-\Trs)\V\A +\Trs
%  -\tfrac{(1-\Trs)\Trs(\V\A-1)^2}{\Trs\V\A + 2-\Trs}\\
%%  &=\tfrac{(2-\Trs)\V\A+\Trs}{\Trs\V\A+2-\Trs}
  =\tfrac{2-\Trs+\Trs/\V\A}{\Trs+(2-\Trs)/\V\A},
  \\
  \entC\E\Y^\het =\log\tfrac{1+1/\V\A}{\Trs+(2-\Trs)/\V\A}
  +\tfrac{(1-\Trs)(1-1/\V\A)}{\Trs+(2-\Trs)/\V\A}
  \log\tfrac{1+1/\V\A}{(1-\Trs)(1-1/\V\A)}.  
\end{gather*}
When $\V\A\gg1/\Trs$, this expression becomes 
\begin{equation*}
  \entC\E\Y^\het =\log\tfrac{1-\Trs}{\Trs}-\f\Trs\log(1-\Trs)
  +\bO{\f{\Trs\V\A}}. 
\end{equation*}
If one also has $\V\A\gg1/(1-\Trs)$, \cite{noteapprox}
\begin{gather*}
  \ent E =\log\V\A + \log \tfrac\e2(1-\Trs)
  +\bO{\f{(1-\Trs)\V\A}} 
  \\
  \I\Y\E^\het =\log \tfrac\e2 \Trs\V\A +\f\Trs\log(1-\Trs)
  +\bO{\{\f\Trs+\f{1-\Trs}\}\f{\V\A}}
  \\
  \cboxed{
    \rDI^\het = \f\Trs\log\f{1-\Trs}- \log\e
    +\bO{\{\f\Trs+\f{1-\Trs}\}\f{\V\A}}
  }
\end{gather*}
As in the direct case, Eve gains a finite amount of information
%\cite{notehetrev} 
by using collective measurement instead of heterodyne measurements.
However this gain is not sufficient to reduce the range of the
protocol, which still works for arbitrary long ranges. For strong
losses ($1/\V\A\ll\Trs\ll1$), the rate is twice smaller than the
collective measurement rate
\begin{equation*}
  \rDI^\het=\f2\Trs\log\e+\bO{\Trs^2+\f{\Trs\V\A}}\simeq\f2\rDI^\coll. 
\end{equation*}

\paragraph{Reverse homodyne key generation}

In the homodyne case \cite{GG-Proc}, similar calculations give us:
\begin{gather*}
  \Vc{\Q[\E]}\Y^\hom=\tfrac1{\Trs+(1-\Trs)/\V\A}\\
  \Vc{\P[\E]}\Y^\hom=\V\E=(1-\Trs)\V\A+\Trs\\
  \Vc\E\Y^\hom=\sqrt{\V\A\tfrac{1-\Trs+\Trs/\V\A}{\Trs+(1-\Trs)/\V\A}}
\end{gather*}
In the large modulation limit ($\V\A\gg 1/\Trs ; 1/(1-\Trs)$)
\cite{noteapprox},
\begin{gather*}
   \Vc\E\Y^\hom
  =\sqrt{\V\A\tfrac{1-\Trs}{\Trs}}
  \left[1+\bO{\{\f\Trs+\f{1-\Trs}\}\f{\V\A}}\right] 
  \\\displaybreak[2] 
  \entC\E\Y^\hom =\f2\log\tfrac{1-\Trs}\Trs\V\A +\log\tfrac\e2
  +\bO{\f{\Trs\V\A} +\sqrt{\tfrac{\Trs}{(1-\Trs)\V\A}}}
  \\
  \I\Y\E^\hom
  =\f2\log(1-\Trs)\Trs\V\A 
  +\bO{\f{\Trs\V\A} +\sqrt{\tfrac{\Trs}{(1-\Trs)\V\A}}}
  \\
  \cboxed{
    \rDI^\hom =\f2\log\f{1-\Trs}
    +\bO{\f{\Trs\V\A} +\sqrt{\tfrac{\Trs}{(1-\Trs)\V\A}}}
    }
\end{gather*}
As for the direct case, the use of homodyne detection by Alice and Bob
allows them to reduce the advantage given to Eve by coherent
measurement to an arbitrarily small amount and to attain a secret key
rate arbitrarily close to the one given in \cite{GG-Proc}, where only
individual attacks were considered. In the
strong losses regime ($1/\V\A\ll\Trs\ll1$), the rate obtained is
almost equal to the one obtained in heterodyne measurements:
\begin{equation*}
  \rDI^\hom=\f2\Trs\log\e+\bO{\Trs^2+\f{\Trs\V\A}}\simeq\rDI^\het.  
\end{equation*}
The (almost) factor 2 advantage in rate given by heterodyne measurement  
at low losses cancels for strong losses.
%
%The achievable rate of reverse homodyne key distribution remains
%however smaller than the one achievable with a heterodyne protocol.
The use of the homodyne setup is also
% remains nevertheless 
attractive in experimental QKD because of the
sensitivity of the current continuous variable reconciliation
algorithms \cite{slice, NVAC} to the signal-to-noise ratio.
 
\paragraph{Unconditional security}

One can also compare these rates to the unconditionaly secure rates
$S$ obtained from the generic security proof of Christiandl \etal\ 
\cite{generic}
\begin{equation}
  \label{eq:generic}
  S=\I\X\Y-\ent\E,
\end{equation}
which is independent of the reconciliation direction. Since Alice
sends coherent states, as shown above $\ent\E=\I\X\E$ and this
unconditional secure rate is equal to the direct reconciliation rate
$\dDI^\coll$ ($\dDI^\het$) when Bob makes collective (heterodyne)
measurements, regardless of the actual reconciliation direction
(direct or reverse).% In the direct case, the optimal joint attack is
%therefore the collective attack considered above. 

The role of the unmeasured ($P$-)quadrature modulation makes the
homodyne case different. If it decreased the efficiency
of the collective attacks considered above, it increases the entropy
$\ent\E$, decreasing the secure rate $S$ given by equation
\eqref{eq:generic}. However, giving the information about this
modulation to Eve could only decrease the secret rate. $\ent\E$ is
then given by \eqref{eq:HBXhom} and, in the strong modulation regime
($\V\A\gg1/(1-\Trs)$),  \eqref{eq:generic} becomes
\cite{noteapprox}
\begin{equation*}
  \cboxed{
    S^\hom
    =\f2\log\tfrac\Trs{1-\Trs}\tfrac4{\e²} 
    +\bO{\f{\sqrt{(1-T)\V\A}}}
  }
\end{equation*}
Unconditionally secure homodyne QKD is therefore possible if the
channel transmission $\Trs$ is greater than $\Trs_{\min}^\hom
=\e²/(\e²+4) \simeq0.65$ (1.9~dB)

\paragraph{Conclusion}

We have quantified the effect of collective attacks on coherent states
based CV QKD protocols through a lossy channel. These attacks are
strictly more powerful than the individual attacks studied before.
However, if Bob makes homodyne measurements, the advantage given to
Eve by these attacks can be made arbitrarily small if Alice uses a
strong modulation. On the contrary, if Bob uses heterodyne
measurement, these attacks give a finite advantage to Eve.  For
comparison, we have computed the key rate in the (theoretical) optimal
case, where Bob performs a collective measurement. 

We also have applied the generic  security proof of
Christiandl \etal\ \cite{generic} to compute an unconditional secure
rate for these protocols. %For the direct
%collective and heterodyne protocols, it gives the rates obtained by
%considering collective attacks. This shows the optimality of these
%attacks. In the other situations, 
This rate is usually lower than the one obtained above, but, this
bound being known not to be tight, this does not rule out the
possibility for the considered collective attacks to be optimal.

\begin{acknowledgments}
  This researsh was supported by a Marie Curie Intra European
  Fellowship within the 6th European Community Framework Programme
  (Contract Number: MEIF-CT-2003-502045). I thank J. I. Cirac for
  enlightening discussions and his hospitality at the MPQ,
  N.~Lütkenhaus for bringing
  \cite{DW} %\cite{DWshort, DWlong} 
  to my attention.
  I acknowledges discussions with Ph. Grangier, M.~M.~Wolf, A.~Acín,
  M.~Navascués, S. Iblisdir,  G. Giedke and B.~Kraus.
\end{acknowledgments}


\begin{thebibliography}{99}
\bibitem{noteapprox} The exact expressions, straightforward to compute
  but not very enlightening, have been omitted.
\end{thebibliography}
\end{document}